\documentclass[11pt]{article}
\usepackage{moriond,epsfig}

\def\Journal#1#2#3#4{{#1} {\bf #2}, #3 (#4)}

\def\be{\begin{equation}}
\def\ee{\end{equation}}
\def\bea{\begin{eqnarray}}
\def\eea{\end{eqnarray}}

\begin{document}
\vspace*{4cm}
\title{ON RADIATION DRAG
AND 
NEUTRONS
IN GRBS}

\author{T. Bulik, M. Sikora, R. Moderski}

\address{Nicolaus Copernicus Astronomical Center, Bartycka 18, 00716 Warsaw,
Poland}

\maketitle

\abstracts{As it was recently shown by Derishev\cite{De}, initial GRB ejecta 
are likely
to contain  comparable number of protons and neutrons.  During acceleration
process, and/or later, due to interaction with external
medium, such ejecta are likely to be split into spatially distinct shells of
neutron and proton-electron plasmas. This leads to dynamical effects
which can affect the afterglow light curves.
In this paper we study these effects including,
for the first time, radiation drag imposed on the neutron rich ejecta.
In the presence of efficient radiation drag and for typical ejecta Lorentz
factor (below 400), this is the neutron shell which moves
faster and, after conversion to proton-electron plasma, powers
the afterglow. After certain amount of deceleration, this shell is hit by
the second one. Such collision will lead to reflaring, as observed in some
afterglow light curves.}

\section{Introduction}

 Recent observational data points toward the
hypernova model  for long gamma-ray bursts. Within the framework of the
hypernova model Lazzati, Ghisellini, Celotti and Rees~\cite{La} show that  the
effects of radiation drag may  be very important for the  dynamics and
radiation of the GRB. The ejecta  encountering   a dense radiation field can be
decelerated and  can produce very hard X-ray spectra by  upscattering the
external radiation field.

Derishev, Kocharovsky, and Kocharovsky~\cite{De} noticed that  the relativistic shock must contain neutrons.  Neutrons
are coupled to protons when the density is high enough, and if the density
becomes low enough already in the acceleration phase then the neutron and
proton ejecta  will separate.

The latter condition can be satisfied only if $\Gamma > 400$. 
In this paper we
point out that the GRB ejecta can split into the neutron and 
the proton-electron shells even for $\Gamma < 400$, 
provided the radiation drag
is efficient like suggested by Lazzati et~al.~\cite{La}.

\begin{figure}
\psfig{figure=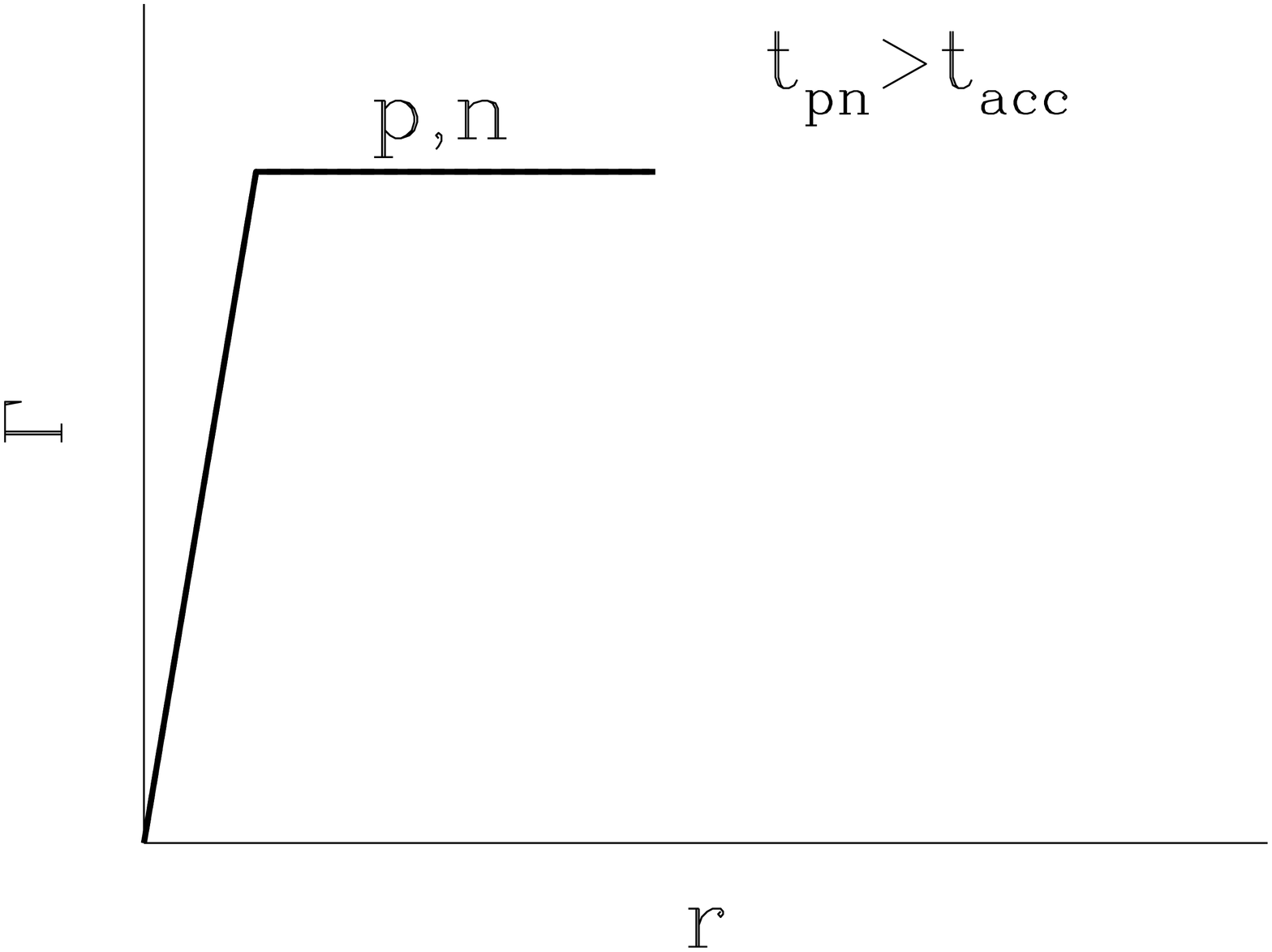,angle=-90,width=0.45\columnwidth}
\psfig{figure=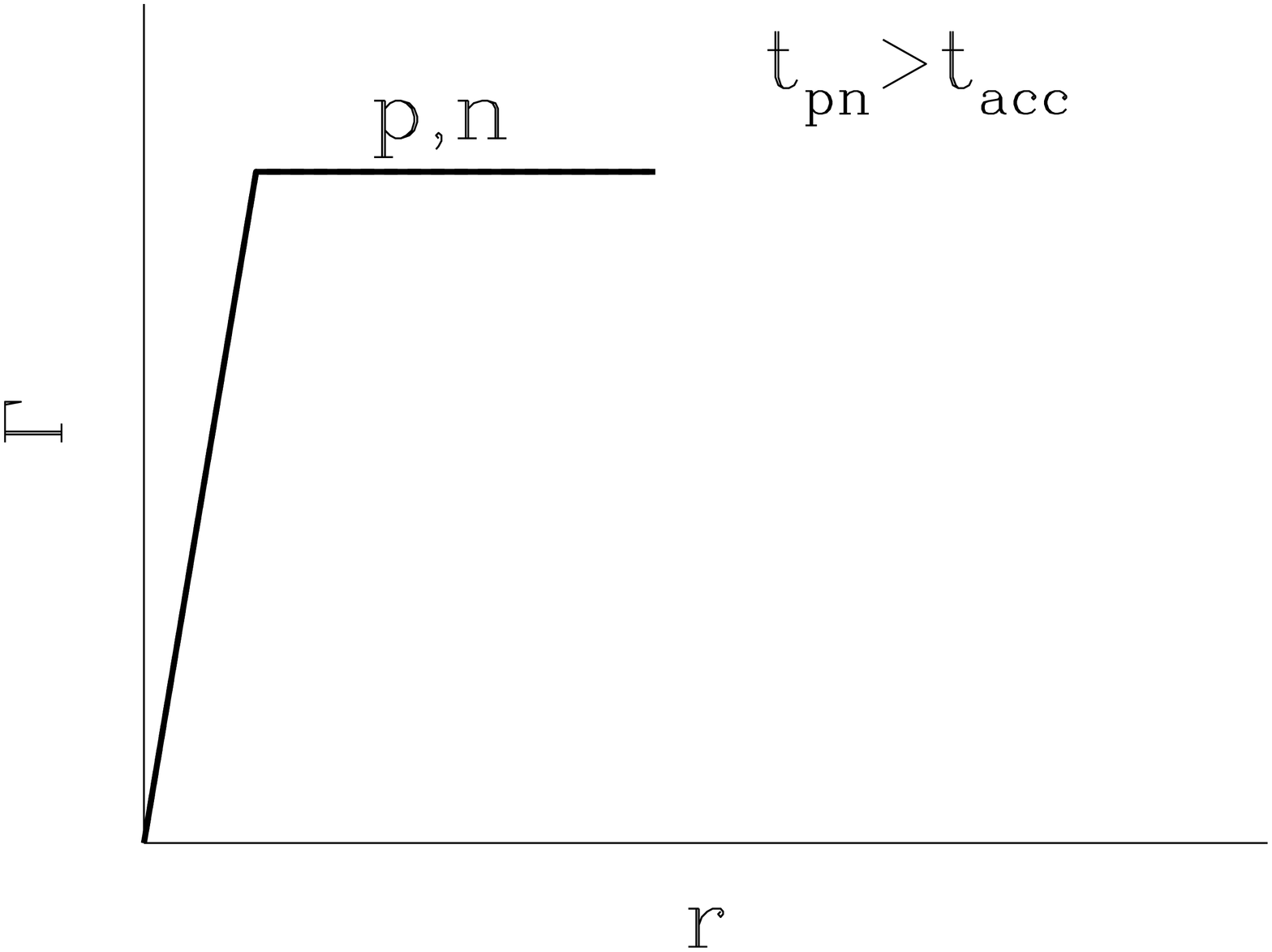,angle=-90,width=0.45\columnwidth}
\caption{The acceleration phase in the presence of neutrons.
If $t_{pn}> t_{acc}$ then the
protons and neutrons move together.
In the opposite case  when $t_{pn}<t_{acc}$
protons and neutrons will separate.}
\end{figure}

\section{The dynamics of the ejecta}

The following timescales, and corresponding distances
 enter the problem:
 
\begin{itemize}
\item $t_{acc}$  - the acceleration timescale;
acceleration ceases at a distance $r_{acc} \simeq c t_{acc}$
 
\item $t_{pn}$ - the time at which the density drops
to the value that
the  proton neutron coupling becomes  inefficient;
the decoupling takes place at $r_{pn}\simeq c t_{pn}$

\item $t_n$  - the neutron decay timescale as measured in the lab frame
($t_n= 900\,{\rm s} \times \Gamma$, where $\Gamma$ is the bulk Lorenz factor
of the ejecta)
 the decay takes place at $\approx r_{n}\simeq c t_n$
 
\item $t_a$ -  beginning of the afterglow; the afterglow starts at a distance
$r_a \simeq c t_a$

\end{itemize}

\subsection{Without radiation drag}
If  $t_{pn}< t_{acc}$  we expect a separation of the
proton and neutron stream. The neutrons are accelerated only because they are
held in the shock by collisions with protons, and once these collisions
become inefficient they lag behind the neutron stream, see Figure 1.
This can take place only if $\Gamma > 400$.

In the afterglow phase two scenarios may take place depending
on  $t_n$ and $t_a$, as shown in Figures 2 and 3.
If $t_n < t_a$
and   $t_{pn}> t_{acc}$ there will be no separation of protons and neutrons, as
shown in the left panel of Figure~2. 
In all other cases neutrons decay to protons
forming  a separate flow which sooner or later collides
with the original proton flow. If $t_n > t_a$ (right panel of Figure~2)
then at the beginning of the afterglow phase only the protons are decelerated,
while the neutrons can stream ahead until they decay to protons at 
$r_n$. At this time they begin decelerating and clear the 
path in front of the original proton shock. The two 
shocks finally collide at $r_{coll}$ when the trailing proton shock 
catches up with the shock formed of the protons formed of the decaying neutrons.

Figure~3 shows the case when the proton and neutron stream is 
separated in the acceleration phase.The two cases shown are very similar:
the afterglow is started by the faster proton stream. The neutrons decay
to protons and the two shocks collide when the initial proton wave has been
slowed down sufficiently so that the trailing - now faster, shock catches up.

\subsection{With radiation drag}
We present the effect of the radiation drag in Figure~4.
 We demonstrate there
the case when $t_{pn}> t_{acc}$, i.e. the protons and the neutrons are
accelerated together and the neutrons decay after the acceleration phase has
ceased. After the acceleration phase the ejecta have to plough through 
the dense radiation field and the protons are decelerated. At this time however
the neutrons are no longer bound to protons, and the two streams separate.
The radiation drag ends at $r_*$ (the radius of the exploding star)
and we now have a leading 
neutron wave followed by the proton ejecta. If the neutron decay radius $r_n$ 
is smaller than the radius $r_a$ at which the afterglow begins (left panel
of Figure~4) then the leading neutrons convert to protons and begin 
decelerating once they reach $r_a$. As they decelerate they sweep up the matter
in front of the trailing proton ejecta. The  two waves collide 
when the leading shock has slowed down sufficiently so that the trailing one 
can catch up, which takes place at $r_{coll}$. The case
when the afterglow radius $r_a$ is smaller than the radius $r_n$ where 
neutrons decay to protons  is shown in the right panel of Figure~4.
The trailing proton shock begins the afterglow,  while the leading 
neutron wave rushes through the external matter without any interaction
until it reaches $r_n$ and converts to protons. At this moment it begins to
decelerate and cleares the way in front of the trailing proton wave.
Once the trailing wave has reached $r_n$ it stops decelerating. 
The two waves finally collide at $r_{coll}$.

\begin{figure}
\psfig{figure=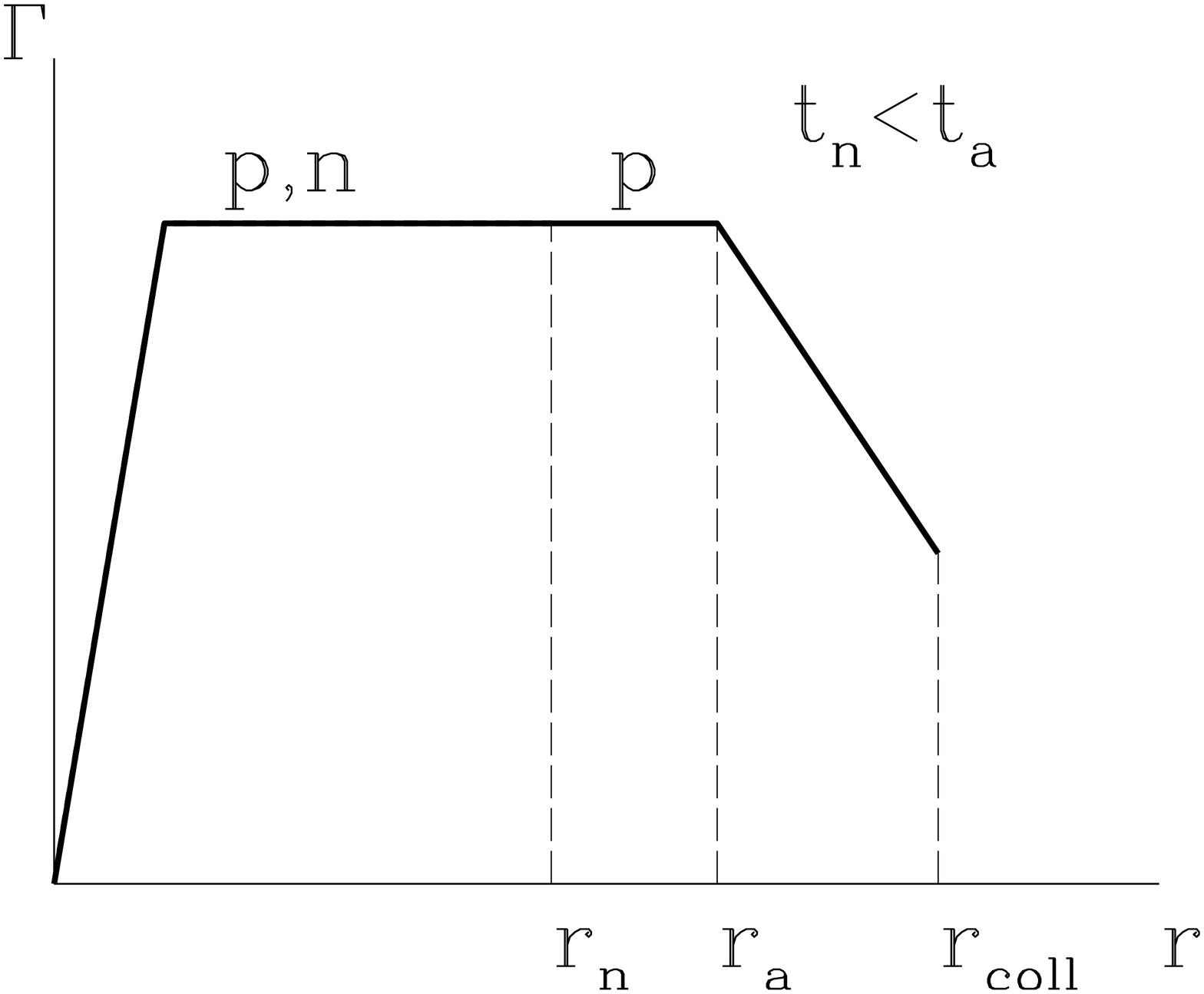,angle=-90,width=0.45\columnwidth}
\psfig{figure=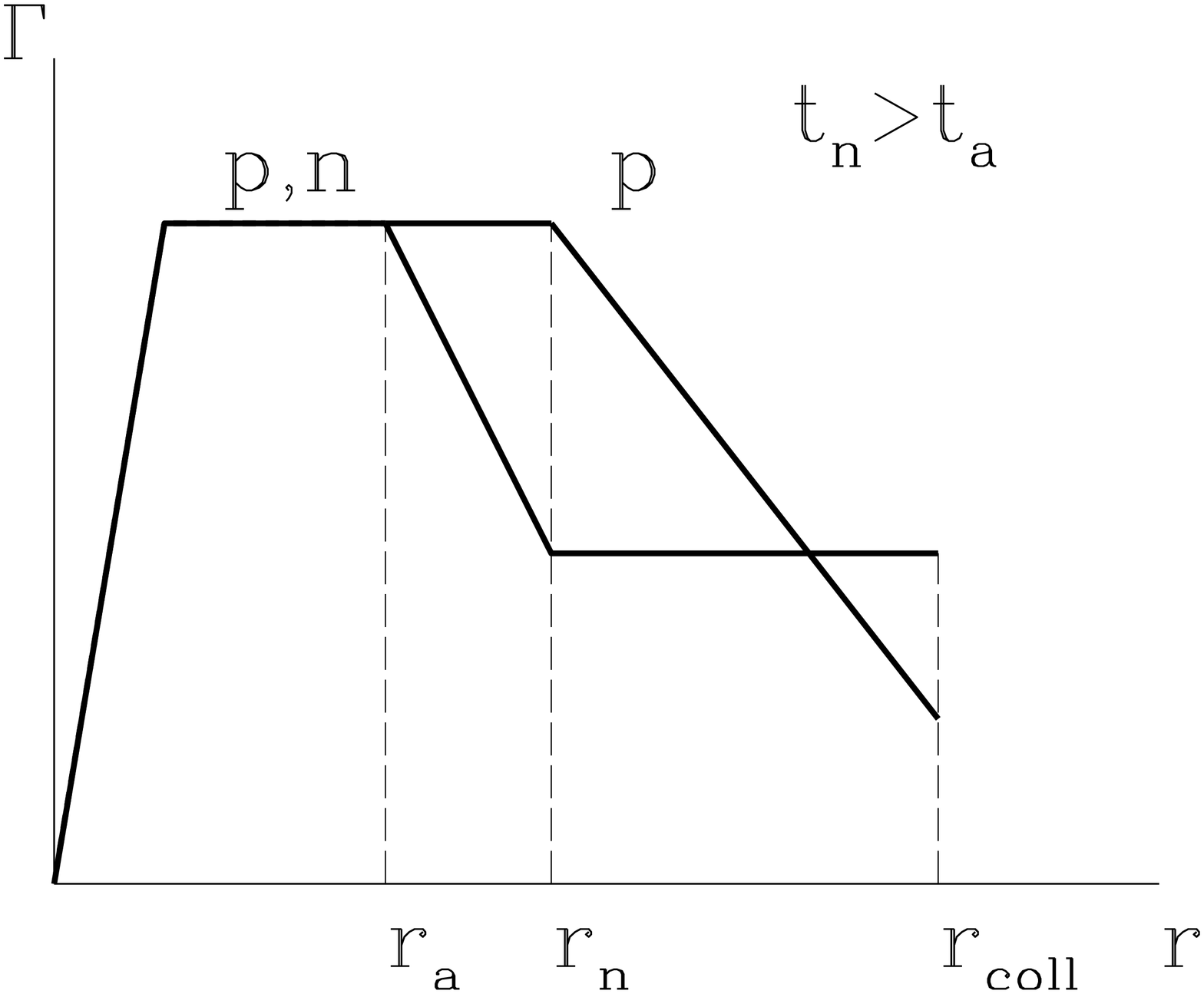,angle=-90,width=0.45\columnwidth}
\caption{The case  $t_{pn}> t_{acc}$: in the left panel we
present the  case, where $t_n < t_a$, and all neutrons decay
to protons before the beginning of the afterglow. There is no separation
of protons and neutrons. In the right panel we present the opposite case,
$t_n > t_a$. Here the afterglow starts at $r_a$ and begins to slow down
protons while the neutrons stream ahead. They decay in front
of the earlier proton afterglow at $r_n$, and sweep up matter in front of the
proton shock.
The protons catch up and collide
with this shock at $r_{coll}$.}
\end{figure}

\begin{figure}
\psfig{figure=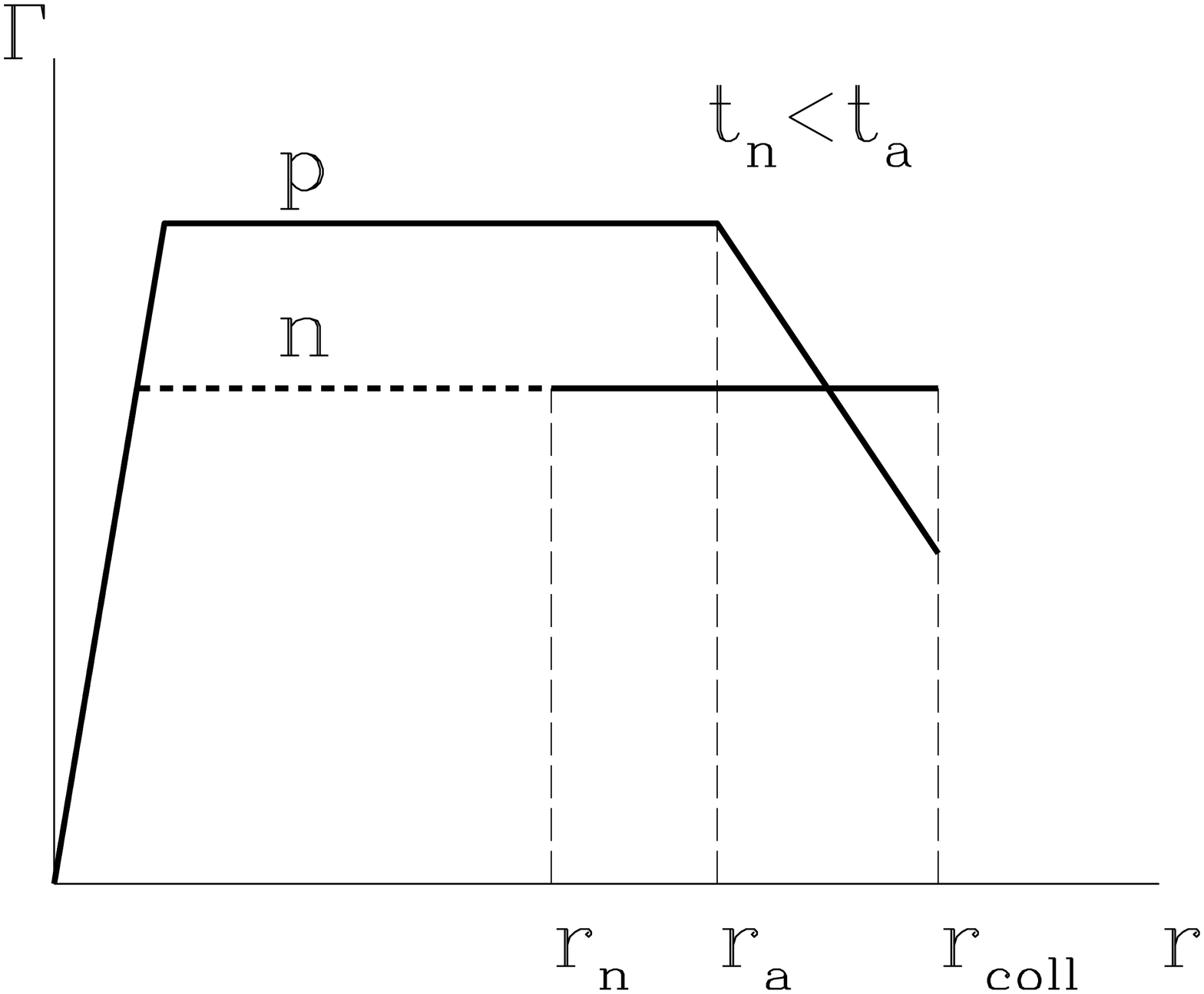,angle=-90,width=0.45\columnwidth}
\psfig{figure=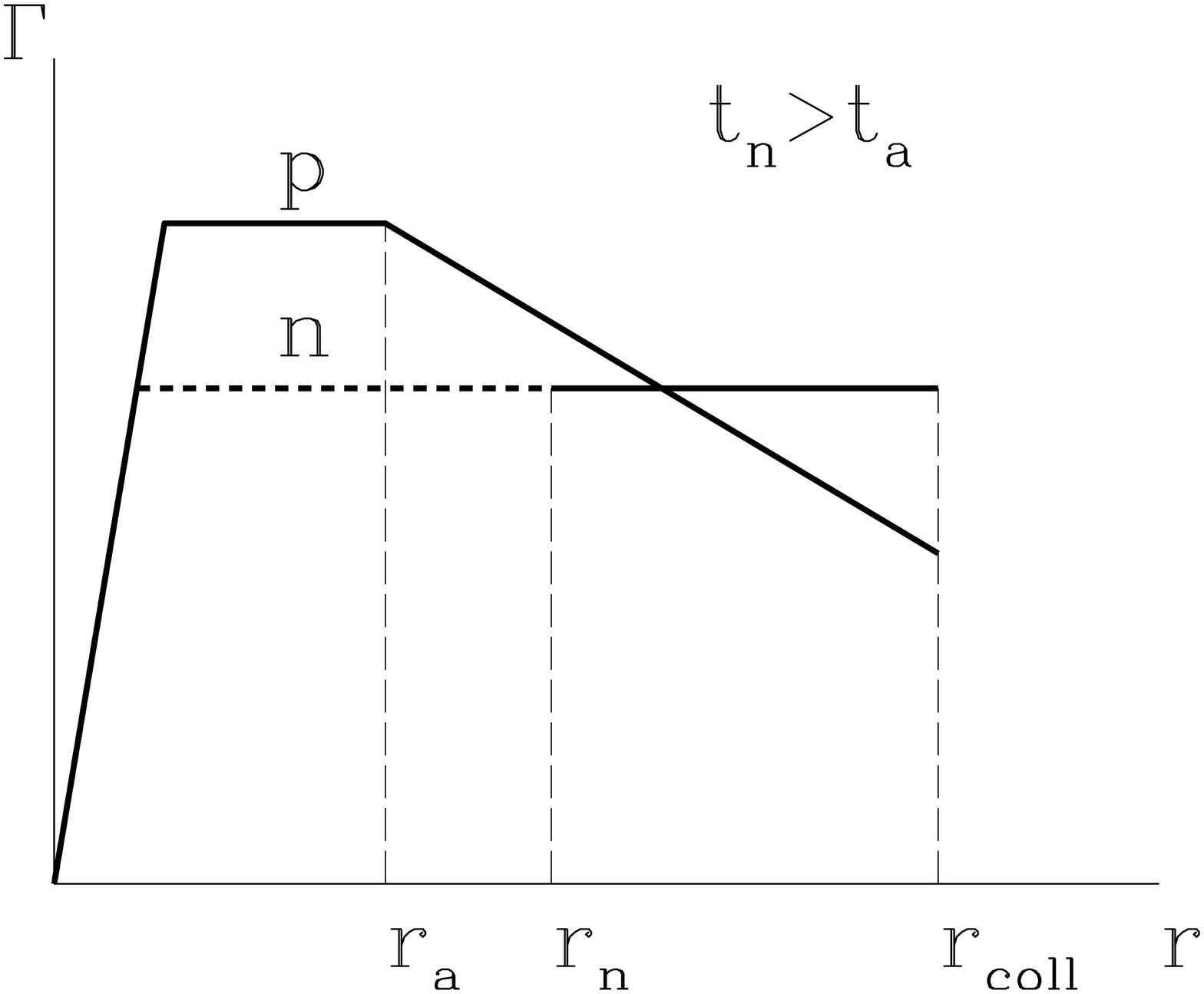,angle=-90,width=0.45\columnwidth}
\caption{The case when $t_{pn}< t_{acc}$. The protons and the
neutrons have been separated already in the acceleration phase.
In the left panel we present the case when $t_n < t_a$.
Neutrons trail behind the protons, and decay still behind them.
Once the afterglow starts and the proton front decelerates the two
shocks have chance to collide. Similarly, when $t_n > t_a$ (right panel),
the trailing neutrons convert to protons and the two shocks collide.
}
\end{figure}

\begin{figure}
\psfig{figure=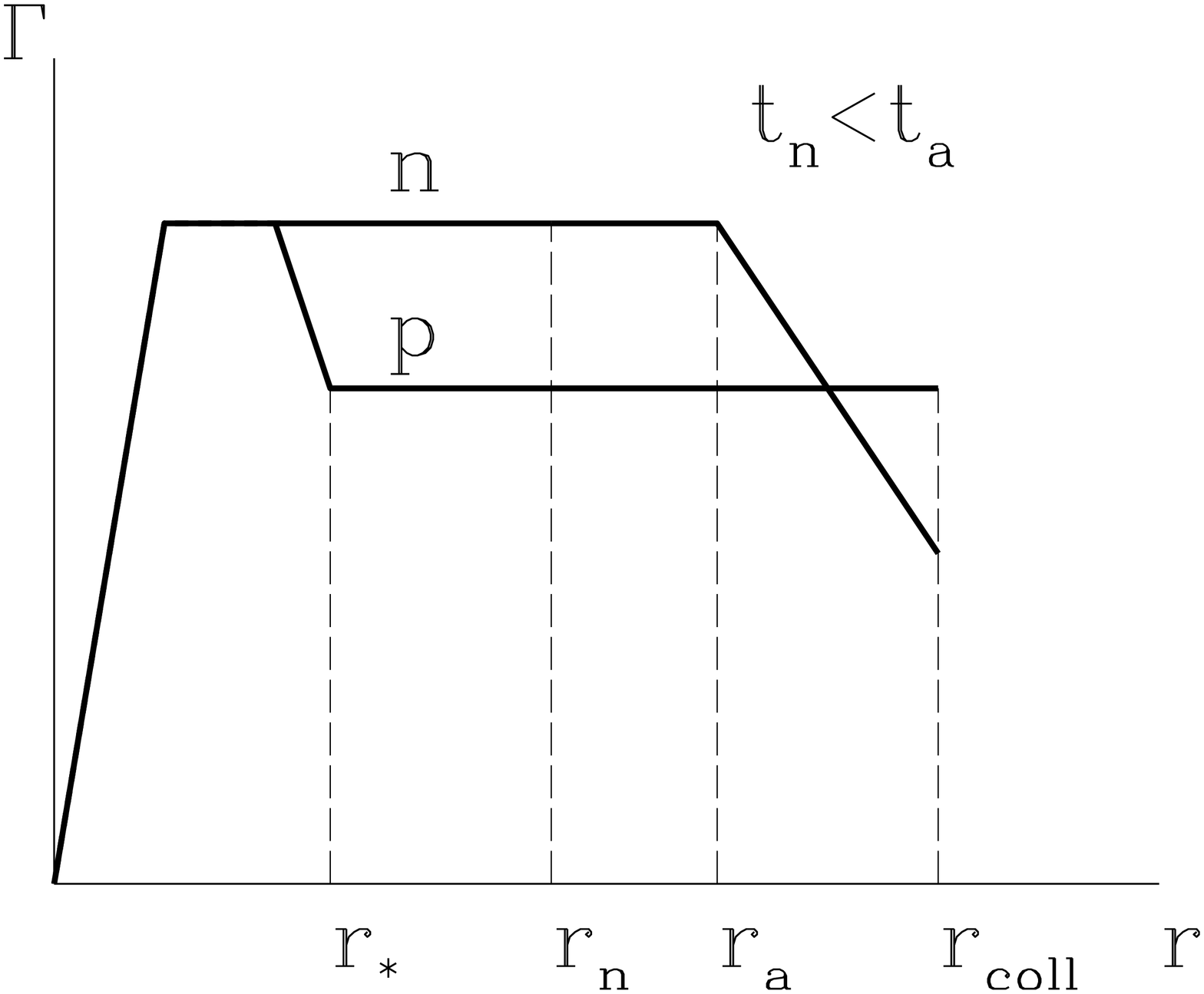,angle=-90,width=0.45\columnwidth}
\psfig{figure=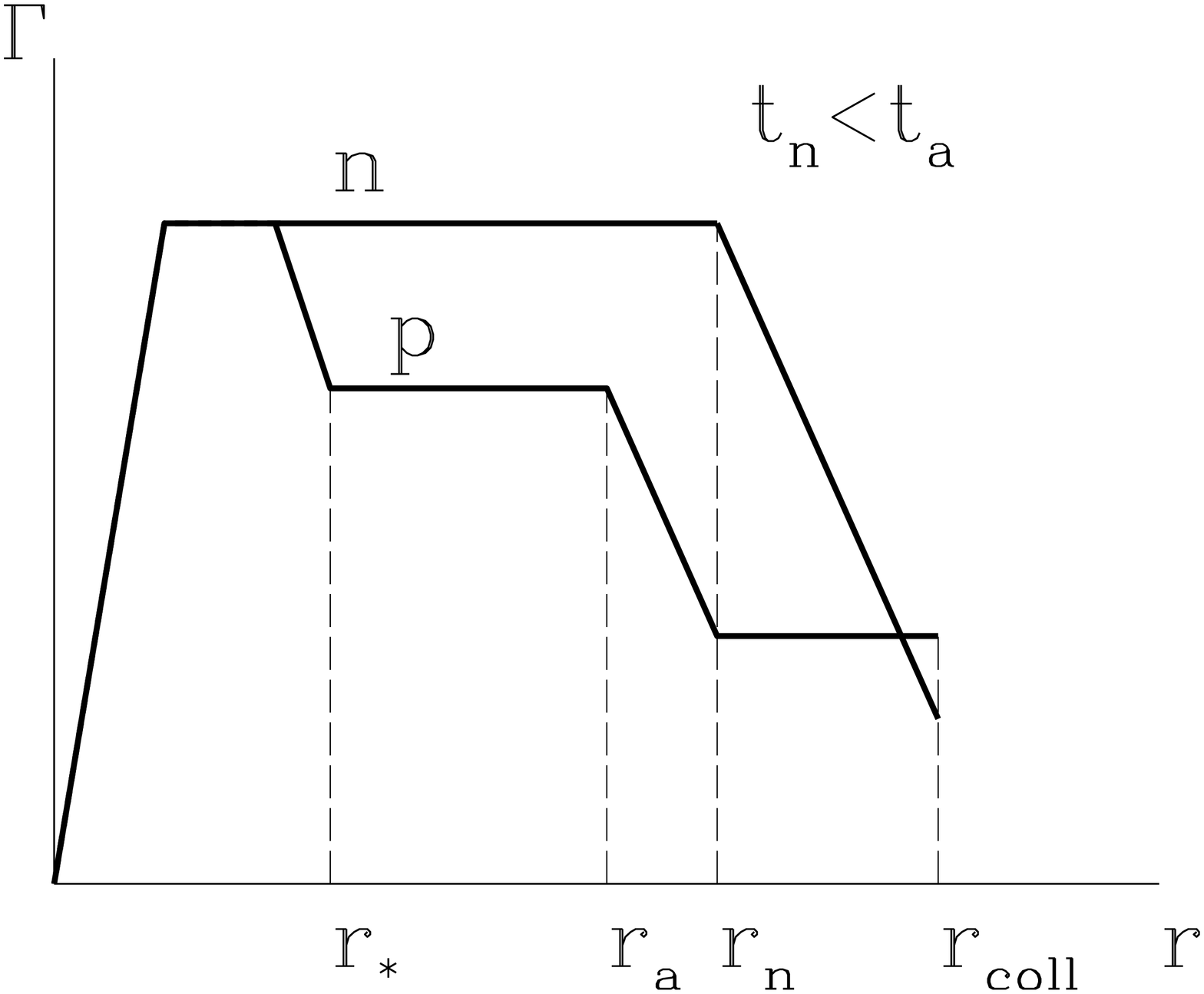,angle=-90,width=0.45\columnwidth}
\caption{The effects of radiation drag. Here we demonstrate
the case $t_{pn}> t_{acc}$. After the acceleration phase ceases
the radiation
forces act on protons and slow them down while the neutrons stream ahead.
We marked the radius where the radiation drag
stops by $r_*$. If  $r_n <r_a$, (left panel)
 neutrons first decay to protons,
starting the afterglow, and the protons catch up with them at $r_{coll}$
(similarly to  top panel in Figure 4).
If  $r_n >r_a$ we encounter a situation similar to the right
panel of Figure~3.
}
\end{figure}

\section{Conclusions}

The inclusion of both the effects of the radiation drag and of the neutron 
content in the ejecta inevitably leads to a collision of
the two shells in the afterglow phase. Such collision will
cause reflaring in  the afterglow.
 
If the external medium is dense enough and
the afterglow starts at the distance shorter than
the distance of the neutron decay,
then in addition to the
reflaring  discussed above the early afterglow will consist
of two components: the first from the proton shell and the second
due to the shock formed when neutrons decay to protons.
This is illustrated in the right panel of Figure~2 
and in the right panel of Figure~4. One component of the afterglow
is emitted between $r_a$ and $r_n$ by the proton wave and the second component 
is formed between $r_n$ and 
$r_{coll}$ by the deceleration of the protons formed in neutron decays.
The two components will very likely overlap in the observers frame.

\section*{Acknowledgments}
We acknowledge support of the KBN grant 
5P03D01120 (TB) and
5P03D00221 (MS and RM).

\end{document}